\documentclass[a4paper]{article}
\usepackage{INTERSPEECH2021}
\usepackage{graphicx}  
\usepackage{float}  
\usepackage{subfigure}  
\usepackage{epsfig}
\usepackage{verbatim}

\title{Adaptive Margin Circle Loss for Speaker Verification}
\name{Runqiu Xiao$^1,^2$,Xiaoxiao Miao$^1,^2$,Wenchao Wang$^1,^2$,Pengyuan Zhang$^1,^2$,Bin Cai$^3$,Liuping Luo$^3$}
\address{
  $^1$ Key Laboratory of Speech Acoustics and Content Understanding, Institute of Acoustics, China\\
  $^2$ University of Chinese Academy of Sciences, Beijing, China\\
  $^3$ Guangdong Provincial Public Security Department}
\email{\{xiaorunqiu,miaoxiaoxiao,wangwenchao,zhangpengyuan\}@hccl.ioa.ac.cn\\jiexiren@gmail.com,jclp1122@163.com}

\begin{document}

\maketitle
\begin{abstract}
Deep-Neural-Network (DNN) based speaker verification systems use the angular softmax loss with margin penalties to enhance the intra-class compactness of speaker embeddings, which achieved remarkable performance. In this paper, we propose a novel angular loss function called adaptive margin circle loss for speaker verification. The stage-based margin and chunk-based margin are applied to improve the angular discrimination of circle loss on the training set. The analysis on gradients shows that, compared with the previous angular loss like Additive Margin Softmax(Am-Softmax), circle loss has flexible optimization and definite convergence status. Experiments are carried out on the Voxceleb and SITW. By applying adaptive margin circle loss, our best system achieves 1.31\%EER on Voxceleb1 and 2.13\% on SITW core-core.

\end{abstract}

\section{Introduction}
The modern Deep Neural Network (DNN) based speaker verification systems have achieved remarkable performances than the traditional i-vector with Probabilistic Linear Discriminant Analysis(PLDA) systems~\cite{dehak2010front,prince2007probabilistic,snyder2018x}. It generally consists of three parts during the training phase, the DNN model for segment-level speaker feature extraction, the pooling layer for statistics extraction, and the loss function for classifying~\cite{snyder2018x,variani2014deep}. In recent years, more and more advanced model architectures are proposed for the improvement in ASV performance. Such as the extensions of the basic TDNN~\cite{snyder2019speaker} structure like TDNN-F~\cite{povey2018semi}, ECAPA-TDNN~\cite{desplanques2020ecapa}, e.g., the primary Residual Network (ResNet)~\cite{cai2018exploring}, and their subsequent like Res2Net, ResNeXt~\cite{zhou2021resnext}, e.g.
\\ \indent In speaker verification, the cross-entropy loss function with softmax is most widely used for training the speaker embedding model. However, the previous work shows that it is more suitable for classification because it only learns features that are not discriminative enough~\cite{liu2017sphereface}, which causes larger generalization errors for unseen speakers. To address this issue, Contrastive Loss~\cite{li2017deep} and Triplet Loss ~\cite{zhang2017end,bredin2017tristounet} were first presented to directly optimize the similarity between the speaker embeddings so that the distance of the intra-class embeddings is smaller than the inter-class embeddings over a threshold. Though they performed well by selecting appropriate training samples, the number of pairs or triplets increases explosively with the number of training samples, and the performance depends strongly on the strategy to search effective pairs or triplets.
\\ \indent On the other hand, some angular-based losses are proposed to boost the discriminative power of face representations in the face recognition field, including SphereFace~\cite{liu2017sphereface},  Am-Softmax~\cite{wang2018additive}, and Arc-Softmax~\cite{deng2019arcface}, which are also resultful for ASV task~\cite{cai2018analysis,liu2019large,zhou2020dynamic,rybicka2020parameter}. To ensure that the embeddings are more distinguishable in the angular direction, they proposed to normalize the weights and features of the classifier~\cite{ranjan2017l2} and add a margin to tighten the decision boundary. However, the angular-based losses have two drawbacks:(1) after normalization, the network will pay more attention to the low-quality samples~\cite{wang2018additive} and may amplify the impact of noisy samples. (2) The performance relys on the super-parameter, which needed to be obtained through brute force search. To address this, ~\cite{zhou2020dynamic,rybicka2020parameter} propose to set the super-parameter according to the cosine angle dynamically.~\cite{deng2020sub} suggest the Sub-center ArcFace for decreasing the influence from noise.
\\ \indent In this paper, we introduce a novel angular-based losses called adaptive margin circle loss~\cite{sun2020circle} for speaker verification. Analysis on gradients shows that circle loss has flexible optimization and definite convergence status when compared with the other angular-based losses. Then we investigate stage-based and chunk-based strategies to generate adaptive margin, which can enhance the intra-class compactness of speaker embeddings. Experiments on VoxCeleb and SITW show that circle loss achieves better performance than the original softmax loss and the common angular-based losses, Am-Softmax and Arc-Softmax.
\section{From original softmax to angular softmax loss }

\subsection{Softmax Loss}
First, we give a brief review of the original softmax loss. The widely used softmax loss is defined as:
\begin{equation}
\begin{aligned}
L_s&=-\frac{1}{N}\sum_{i=1}^{N}{{\rm{log}}\frac{e^{\textbf w_{y_i}^{T}\textbf x_i + b_{y_i}}}{\sum_{j=1}^{C}{e^{\textbf w_{j}^{T}\textbf x_i + b_j}}}}\\
&=-\frac{1}{N}\sum_{i=1}^{N}{{\rm log}\frac{e^{\left\|\textbf w_{y_i}\right\| \cdot \left\|\textbf x_i\right\|\cdot cos\theta_{y_i}}}{\sum_{j=1}^{C}{e^{\left\|\textbf w_{j}\right\| \cdot \left\|\textbf x_i\right\|\cdot cos\theta_{j}}}}}
\end{aligned}
\label{eq1}
\end{equation}
where C and N is the number of speakers and samples in the mini batch, respectively. $\textbf x_i$ is the input of the last classify layer and $\textbf w_j$ is the $j$-th column of the weights in the classify layer, $\textbf y_i$ is the ground truth label for the $i$-th sample. For convenience, we omit the bias $b_j$ and the logit $\textbf w_{y_i}^{T}\textbf x_i$ is equivalent to $\left \|\textbf w_{y_i}\right \|\left \|\textbf x_i\right \|cos\theta_{y_i}$, where $\theta_{y_i}$ is the angle between $\textbf w_{y_i}$ and $\textbf x_i$. In order to reduce $L_s$, the network tends to:
\begin{itemize}
\item Increase the weight norm $\left \|\textbf w_{y_i}\right \|$. So the more training samples in the i-th class, the larger the corresponding weight norm tends to be~\cite{liu2017sphereface}.
\item Increase the feature norm $\left \|\textbf x_i\right \|$, making the simple samples have greater feature norm~\cite{ranjan2017l2}.
\item Decrease $\theta_{y_i}$. Assuming that $\theta_{y_i}$ belongs to $[0, \frac{\pi}{2}]$.
\end{itemize}
However, the weight norm $\left \|\textbf w_{y_i}\right \|$ and the feature norm $\left \|\textbf x_i\right \|$ are generally useless in the open-set recognition problem. So the authors ~\cite{wang2018additive,deng2019arcface} proposed to normalize the weight and feature vectors(making $\left \|\textbf w_{y_i}\right \|$=$\left \|\textbf x_i\right \|$=1), ensuring that the embeddings $\textbf x_i$ are more distinguishable in the angular direction.

\subsection{Angular Softmax Loss}
The general formula of the Angular Softmax Loss function can be summarized as:
\begin{equation}
\begin{aligned}
L_{as}&=-\frac{1}{N}\sum_{i=1}^{N}{\rm log}\frac{{e^{s\cdot \psi(\theta_{y_i})}}}{{e^{s\cdot \psi(\theta_{y_i})}} + \sum_{j=1,j\ne i}^{C}{e^{s\cdot cos(\theta_j)}}} \\
\psi(\theta_{y_i})&=cos(m_1 \theta_{y_i} + m_2)-m_3
\end{aligned}
\label{eq1}
\end{equation}
where $s$ is the scale factor that makes the lower bound of $L_{as}$  close to 0. The $m1$, $m2$ and $m3$ are used to tighten the decision boundary. When $m1$, $m2$, and $m3$ are used individually, the losses are denoted as angular softmax (A-Softmax), additive angular margin softmax (Arc-Softmax), and additive margin softmax loss (Am-Softmax), respectively.
\\ \indent Without loss of generality, we analyze the gradients of Am-Softmax under the toy scenario where there are only a single $s_p$ and $s_n$:
\begin{equation}
\begin{aligned}
L_{am-s}=-log\frac{{e^{s\cdot (s_p - m)}}}{{e^{s\cdot (s_p - m)}} + (C-1)e^{s\cdot s_n}}
\end{aligned}
\label{eq3}
\end{equation}
where $s_p=cos\theta_{y_i}$ and $(C-1)e^{s\cdot s_n}=\sum_{j=1,j\ne i}^{C}{e^{s\cdot cos(\theta_j)}}$. $s_p$ and $s_n$ refer to the  positive pairs similarity and negative pairs similarity, notice that generally, both $s_p$ and $s_n$ belong to $[0,1]$. The gradients of $L_{am-s}$ with respect to $s_p$ and $s_n$ are derived as follows:
\begin{equation}
\begin{aligned}
\frac {\partial L}{\partial s_p}&=(1-\frac{{e^{s\cdot (s_p - s_n - m)}}}{{e^{s\cdot (s_p - s_n - m)}} + (C-1)})\cdot s\\
\frac {\partial L}{\partial s_n}&=(\frac{(C-1)}{{e^{s\cdot (s_p - s_n - m)}} + (C-1)}) \cdot s
\end{aligned}
\label{eq1}
\end{equation}
As shown in Figure 1(a), the gradients with both $s_p$, $s_n$ are the same to each other and only depend on ($s_p - s_n$). For some noisy training samples, if $s_p$ is small and $s_n$ already close to 0(such as A(0.2, 0.2)), both $s_p$ and $s_n$ still get a large gradient, which means that the loss function keeps on penalizing $s_n$ with a large gradient though $s_n$ has reach optimum. So angular softmax loss will amplify the impact of noise samples.
\section{Adaptive Margin Circle Loss}
\subsection{Circle loss}
In [22], the author provided a self-paced weighting strategy to enhance the optimization flexibility. The proposed Circle loss is defined as
\begin{equation}
\begin{aligned}
L_{circle}&=-{\rm log}\frac{e^{s\cdot \alpha_p \cdot(s_p-\Delta_p)}}{e^{s\cdot \alpha_p \cdot(s_p-\Delta_p)} + \sum_{j=1,j\ne i}^{C}e^{s\cdot \alpha^j_n \cdot(s^j_n-\Delta_n)}} \\
\alpha_p &= O_p-s_p, \quad \alpha^j_n = s^j_n+O_n
\end{aligned}
\label{eq1}
\end{equation}
where $\alpha_p$, $\alpha_n$ are the self-paced weight, and $O_p$, $O_n$ are the optimum for $s_p$, $s_n$, respectively. $\Delta_p$ and $\Delta_n$ are the between-class and within-class margins. When $s_p$ deviates far from $O_p$ or $s_n$ deviates far from $O_n$, $s_p$ or $s_n$ will get effective update with large gradient. If $s_p$ or $s_n$ has reached its optimum, they will get no gradient update. Then the authors proposed to reduce the hyper-parameters by setting $O_p=1+m$, $O_n$=-m, $\Delta_p$=1-m, $\Delta_n$=m. Consequently, the circle loss finally becomes:
\begin{equation}
\begin{aligned}
L_{circle}&=-{\rm log}\frac{e^{s\cdot (m^2-(1-s_p)^2)}}{e^{s\cdot (m^2-(1-s_p)^2)} + \sum_{j=1,j\ne i}^{C}e^{s\cdot ((s^j_n)^2-m^2)}}
\end{aligned}
\label{eq1}
\end{equation}
where the decision boundary is $(1-s_p)^2+s_n^2=2m^2$, the arc of a circle so that the loss function is referred as circle loss. It aims to optimize $s_p$ to 1 and $s_n$ to 0. Similar to the assumption of equation(3), the gradients of circle loss under the toy scenario can be denoted as:
\begin{equation}
\begin{aligned}
\frac {\partial L}{\partial s_p}&=(1-\frac{{e^{s\cdot (2m^2-(1-s_p)^2-s_n^2)}}}{{e^{s\cdot (2m^2-(1-s_p)^2-s_n^2)}} + (C-1)})\cdot 2s \cdot (1-s_p)\\
\frac {\partial L}{\partial s_n}&=(\frac{(C-1)}{{e^{s\cdot (2m^2-(1-s_p)^2-s_n^2))}} + (C-1)}) \cdot 2s \cdot s_n
\end{aligned}
\label{eq1}
\end{equation}
\begin{figure}[!htbp] 
	\centering  
	\vspace{-0.35cm} 
	\hspace{-0.35cm} 
	\subfigtopskip=2pt 
	\subfigbottomskip=2pt 
	\subfigcapskip=-2pt 
	\subfigure[Am-Softmax Loss]{
		\includegraphics[width=0.95\linewidth]{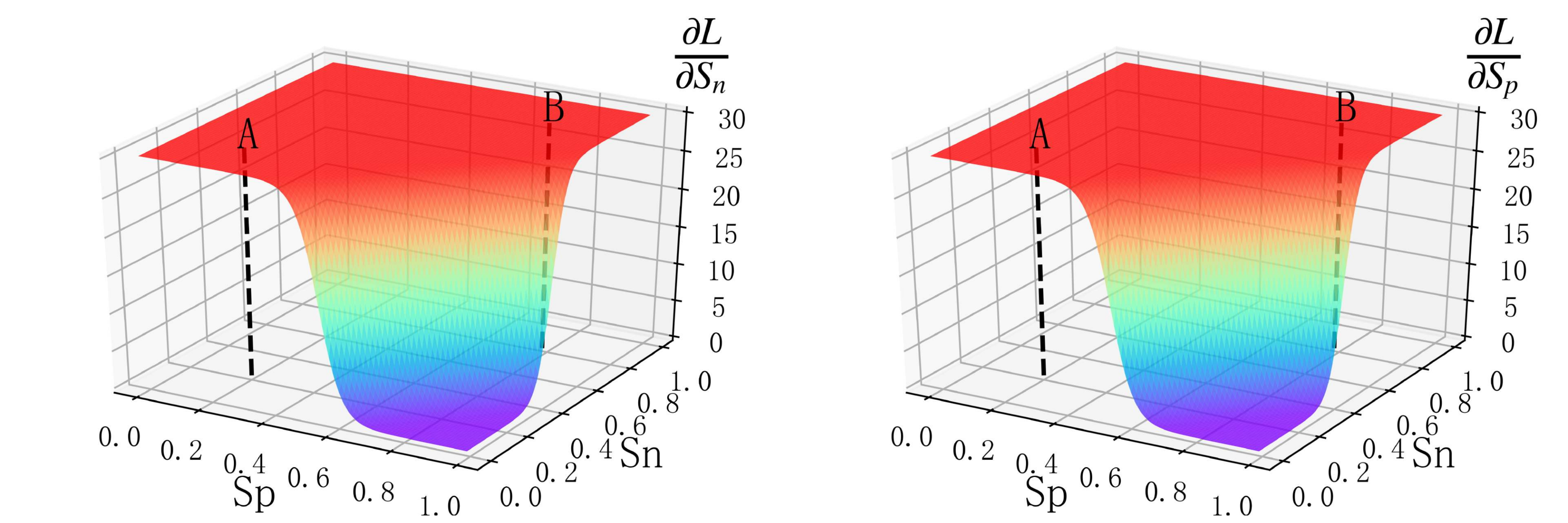}}
    \subfigure[Circle Loss(m=0.40)]{
		\includegraphics[width=0.95\linewidth]{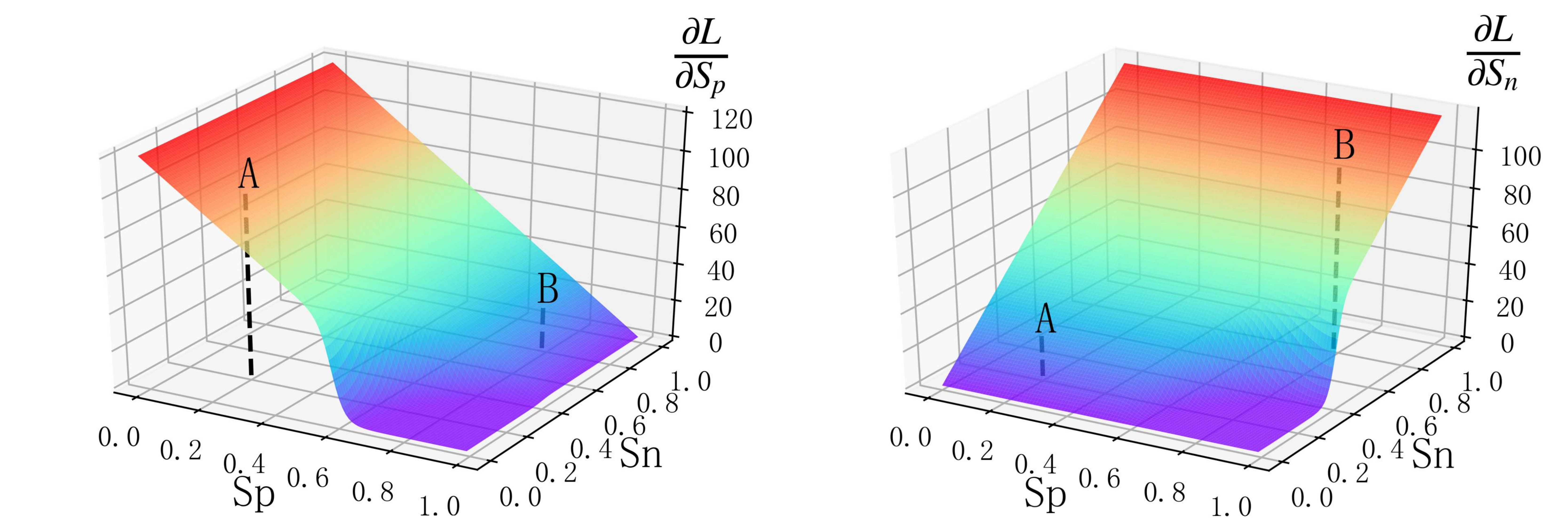}}
    \subfigure[Circle Loss(m=0.35)]{
		\includegraphics[width=0.95\linewidth]{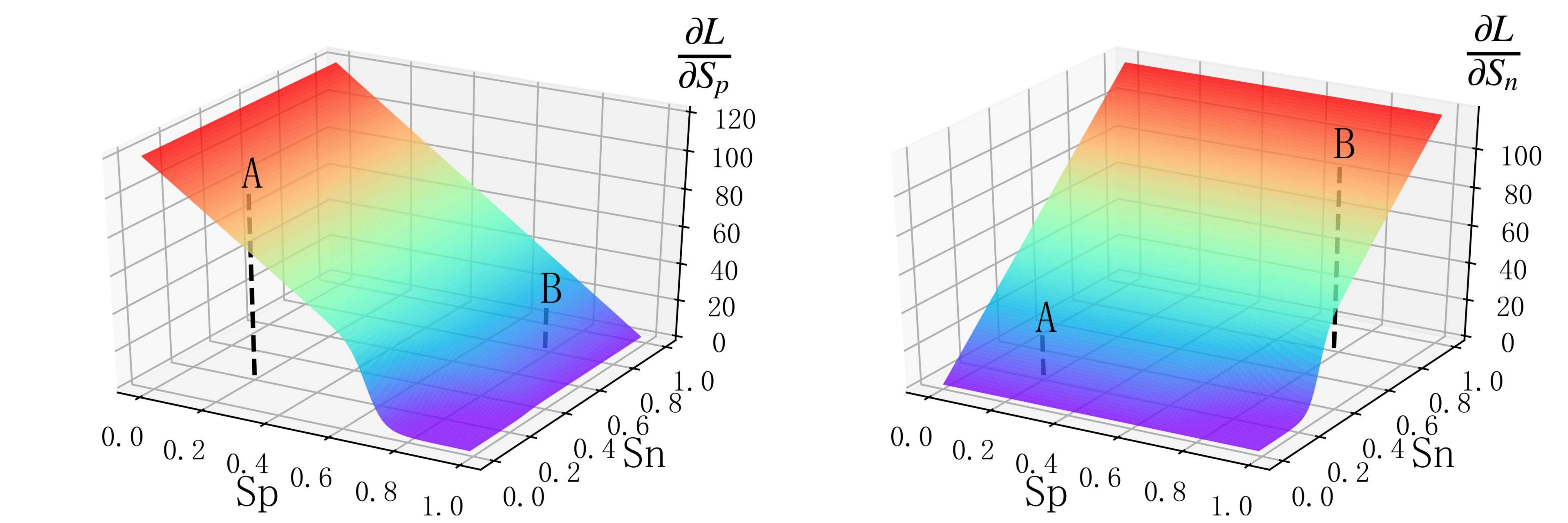}}
    \subfigure[Circle Loss(m=0.25)]{
		\includegraphics[width=0.95\linewidth]{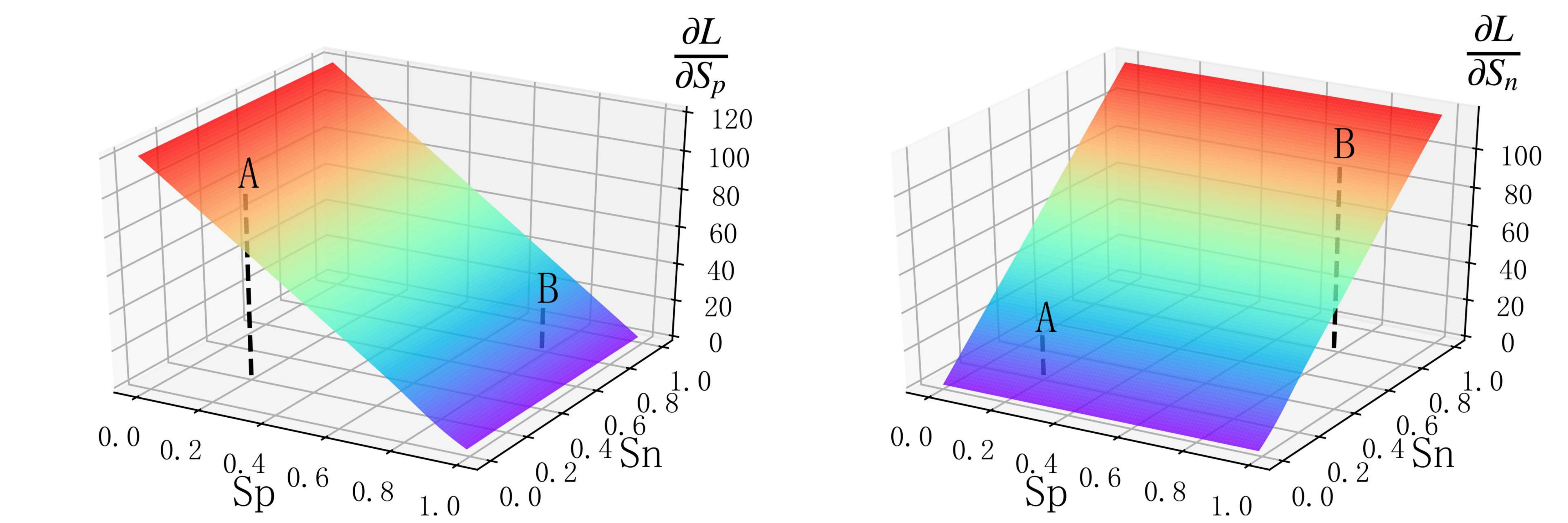}}
	\caption{The gradients of the loss functions. (a)Am-Softmax. (b,c,d) Circle loss with different margin.}
\end{figure}

We visualize the gradients of circle loss with different margin in Figure 1, from which we obtain the following observations:
\begin{itemize}
\item Compared with Am-Softmax, circle loss gives gradually attenuated gradients on $s_p$ and $s_n$. As they gradually reach their optimal, the gradients correspondingly decay, reducing the influence of noisy samples. For instance, point A(0.2, 0.2) gets larger gradients on $s_p$ and smaller gradients on $s_n$, on the contrary point B(0.8, 0.8) gets larger gradients on $s_n$ and smaller gradients on $s_p$.
\item In Figure 1(d), if the margin is too small, the gradient will degenerate into a linear function, and the loss function keeps on penalizing both $s_p$ and $s_n$. In Figure 1(b), a larger margin allows $s_p$ and $s_n$ to converge easily, but the gradient quickly approaches 0 if $s_p$ and $s_n$ cross the decision boundary, which means the loss function will not optimize $s_p$ and $s_n$.
\end{itemize}

Based on the above analysis, we investigate two strategies to generate an appropriate margin that balances convergence speed and discrimination of $s_p$ and $s_n$. 
\subsection{Stage-based margin}
After training with fixed margin circle loss, we randomly sample 10\% training samples and calculate the mean of $s_{p-mean}$ and $s_{n-mean}$ of them. The mean radius for each epoch is defined as: $r_{mean}=\sqrt{(1-s_{p-mean})^2+(s_{n-mean})^2}$. As shown in Figure 2 and Table1, with a lower margin $m$, such as $m=0.25$, it has a lower mean radius and better angle discrimination in the training set. However, its performance is much worse than the system with $m=0.40$. With the unreachable decision boundary, the circle loss keeps on optimizing $s_p$ and $s_n$ and eventually causes the model to overfit the noisy samples. So that we propose the stage-based margin for circle loss, which initializes a larger margin for $m$ and decreases it in the different training stages. The model can converge quickly and reasonably in the first training stage under loose constraints, and the constraints on the model become stricter when the model has learned identification information, making the model have better angular discrimination when converging.

\begin{figure}[htbp]
  \centering
  \includegraphics[width=0.9\linewidth]{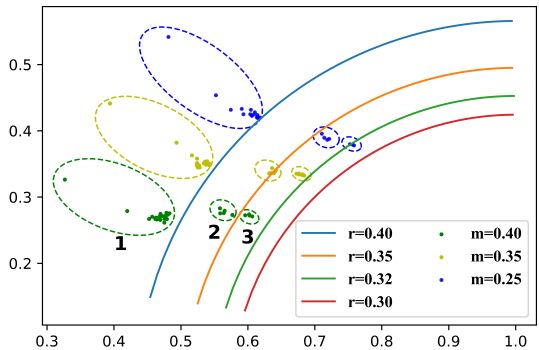}
  \caption{The change of mean radius during different training stage with fixed margin circle loss. The number 1, 2 and 3 represent three stages, from left to right.}
  \label{fig:speech_production}
\end{figure}

\subsection{Chunk-based margin}
Inspired by~\cite{meng2021magface}, the authors propose an adaptive mechanism based on the magnitude which can measure the quality of the given sample. This prevents models from overfitting on noisy low-quality samples. In ASV tasks, we generally randomly cropped or extended the training sample to L frames, while L is randomly sampled from the interval $[L_{min}, L_{max}]$. The training sample is harder for the model when L is smaller, so we propose an adaptive margin based on the chunk width. The formulation is:
\begin{equation}
\begin{aligned}
m=(1-\lambda\frac{L-L_{min}}{L_{max} - L_{min}})m_0
\end{aligned}
\label{eq1}
\end{equation}
where $\lambda$ is hyper-parameters and $m_0$ is the original margin. Parameter $\lambda$ controls the effect of chunk width on the margin. When L is close to $L_{max}$, the margin is smaller and the constraints on the model become stricter.

\section{Experimental setup}
\subsection{Dataset}
We run experiments on the VoxCeleb dataset~\cite{chung2018voxceleb2, nagrani2017voxceleb} and SITW~\cite{mclaren2016speakers}. The development set of VoxCeleb2 (5994 speakers) is used for training. The whole VoxCeleb1 and SITW are used as the evaluation set with four publicly available test trails: VoxCeleb1-O-clean, VoxCeleb1-E-clean, VoxCeleb1-H-clean, and core-core from SITW. 
\\ \indent The equal error rate (EER) and minimum detection cost function with $P_{target}$ equal to 0.01 are presented to demonstrate the performance.

\subsection{Training details}
In our experiments, the input features are the 64-dimensional log Mel-filterbank energies with cepstral mean and variance normalization. No voice active detection (VAD) or data augmentation is applied to the training data. 
\\ \indent Following the previous work in~\cite{cai2018exploring, wang2020investigation},  we use the standard ResNet-34 architecture to extract speaker embeddings. The initial number of channels is set to 32, and only the mean of the frame-level features is used as statistics. Besides, no dropout is applied in our networks.
\\ \indent All models are trained using the stochastic gradient descent (SGD) optimizer with momentum 0.9 and weight decay 1e-3. The learning rate is started with 0.1 and is reduced by 10x when the training loss reaches stability. Mini-batch size is set to 64.
\\ \indent For every training step, a chunk-width L is randomly sampled from the interval $[\rm L_1, \rm L_2]$ and each training sample in the mini-batch is randomly cropped or extended to L frames. We allow $\rm L_1$ and $\rm L_2$ to increase when the learning rate drops. 
According to~\cite{li2019towards},  with a large initial learning rate, the model learns hard-to-generalize, easily fit patterns. So we give the model fewer features by applying a smaller chunk-width L to prevent overfitting in the first training stage. The finally used interval is set to [200,400], [300,500], and [400,600] in the three training stages.
\\ \indent When training with the angular sofmtax loss, we follow the best set in~\cite{liu2019large}, $m=0.2$ and $s=30$ for Am-Softmax, $m=0.25$ and $s=30$ for Arc-Softmax. As to circle loss, $s$ is set to 60. The cosine similarity is used as the back-end scoring method.

\begin{table*}[htbp]
  \centering
  \caption{The results on the VoxCeleb1 test set, the extended and hard test sets (VoxCeleb1-E and VoxCeleb1-H, repsectively), and the evaluation set of SITW Core. The cleaned trial lists are used for Voxceleb.}
    \begin{tabular}{ccccccccc}
    \toprule
    \textbf{Loss} &        & \multicolumn{2}{c}{\textbf{VoxCeleb1-O}} & \multicolumn{2}{c}{\textbf{VoxCeleb1-E}} & \multicolumn{2}{c}{\textbf{VoxCeleb1-H}} & \textbf{SITW Core} \\
    \midrule
           &        & \textbf{EER(\%)} & \textbf{MinDCF} & \textbf{EER(\%)} & \textbf{MinDCF} & \textbf{EER(\%)} & \textbf{MinDCF} & \textbf{EER(\%)} \\
    \midrule
    Softmax &        & 1.77   & 0.192  & 1.79   & 0.202  & 3.23   & 0.306  & 3.25 \\
    \midrule
    Arc-Softmax & s=30, m=0.25 & 1.64   & 0.170  & 1.63   & 0.177  & 2.91   & 0.273  & 2.52 \\
    \midrule
    Am-Softmax & s=30, m=0.20 & 1.71   & 0.161  & 1.65   & 0.183  & 2.83   & 0.263  & 2.46 \\
    \midrule
    \midrule
    Circle & s=60, m=0.25 & 1.68   & 0.200  & 1.74   & 0.194  & 2.96   & 0.275  & 2.46 \\
           & s=60, m=0.30 & 1.75   & 0.160  & 1.74   & 0.189  & 3.02   & 0.286  & 2.52 \\
           & s=60, m=0.35 & 1.44   & 0.161  & 1.58   & 0.170  & 2.72   & 0.258  & 2.36 \\
           & s=60, m=0.40 & 1.45   & \textbf{0.133 } & 1.56   & 0.166  & 2.64   & \textbf{0.240 } & 2.27 \\
    \midrule
    Circle-Chunk & m: 0.40 & 1.41   & 0.145  & 1.55   & 0.162  & 2.67   & 0.253  & 2.19 \\
    \midrule
    Circle-Stage & m: 0.40, 0.35, 0.30 & 1.35   & 0.144  & 1.54   & 0.165  & 2.65   & 0.251  & 2.24 \\
           & m: 0.40, 0.35, 0.32 & \textbf{1.31} & 0.135  & \textbf{1.51 } & \textbf{0.163 } & \textbf{2.61} & 0.250  & \textbf{2.13} \\
    \bottomrule
    \end{tabular}%
  \label{tab:addlabel}%
\end{table*}%
\section{Results}
Table 1 summarizes the results with different loss functions. The first row shows the performance of our baseline system, which is consistent with the results reported in~\cite{wang2020investigation}. Compared with softmax loss, the second and third row shows that ArcSoftmax and Am-Softmax achieve a similar improvement in all test sets. The performance of the circle loss is exhibited in the following sections of Table 1. 
\\ \indent We investigate the influence of different margins for circle loss in the fourth row.  It is clear that the larger the margin $m$, the better the performance of circle loss. And it achieves the best result when $m=0.40$, which reduces $11.6\%$ and $21.8\%$ in terms of EER and MinDCF in Voxceleb1-O compared with Arc-Softmax. Besides, there is a performance gap between $m=0.40$ and $m=0.25$. As shown in Figure 2, it is helpful for circle loss to get better performance by training with the reachable decision boundary.

\begin{figure}[htbp] 
    \setlength{\abovecaptionskip}{-0.cm}
    \setlength{\belowcaptionskip}{-1cm}
	\centering  
	\vspace{-0.cm} 
	\hspace{-0.cm} 
	\subfigtopskip=5pt 
	\subfigbottomskip=5pt 
	\subfigcapskip=5pt 
	\subfigure[The change of $s_p$ and $s_n$ values during the three training stage.]{
		\label{level.sub.1}
		\includegraphics[width=8cm, height=6cm]{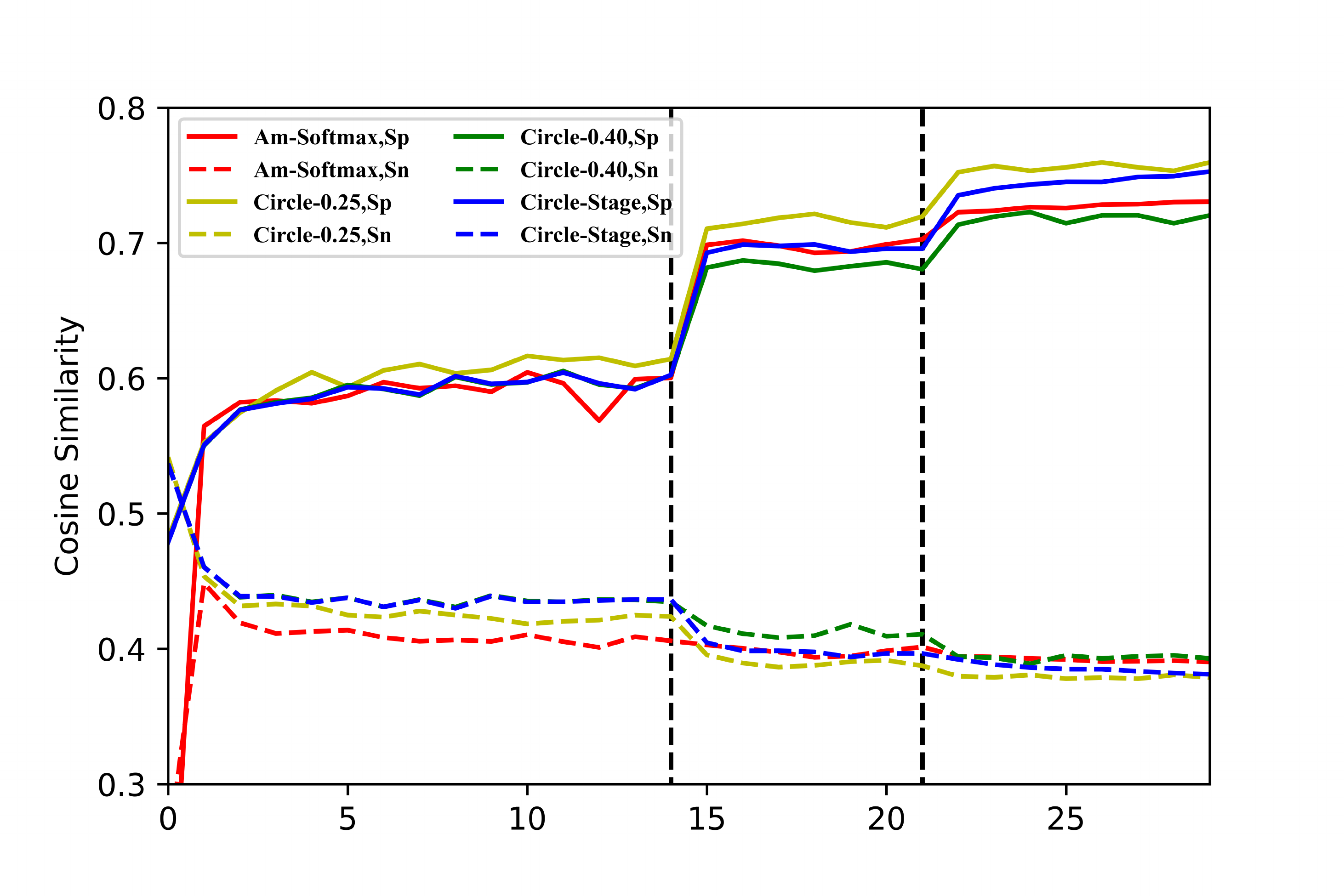}}\\
	\subfigure[The cosine similarity distributions for the same and different speaker spaces in test set.]{
		\label{level.sub.2}
		\includegraphics[width=6cm, height=4cm]{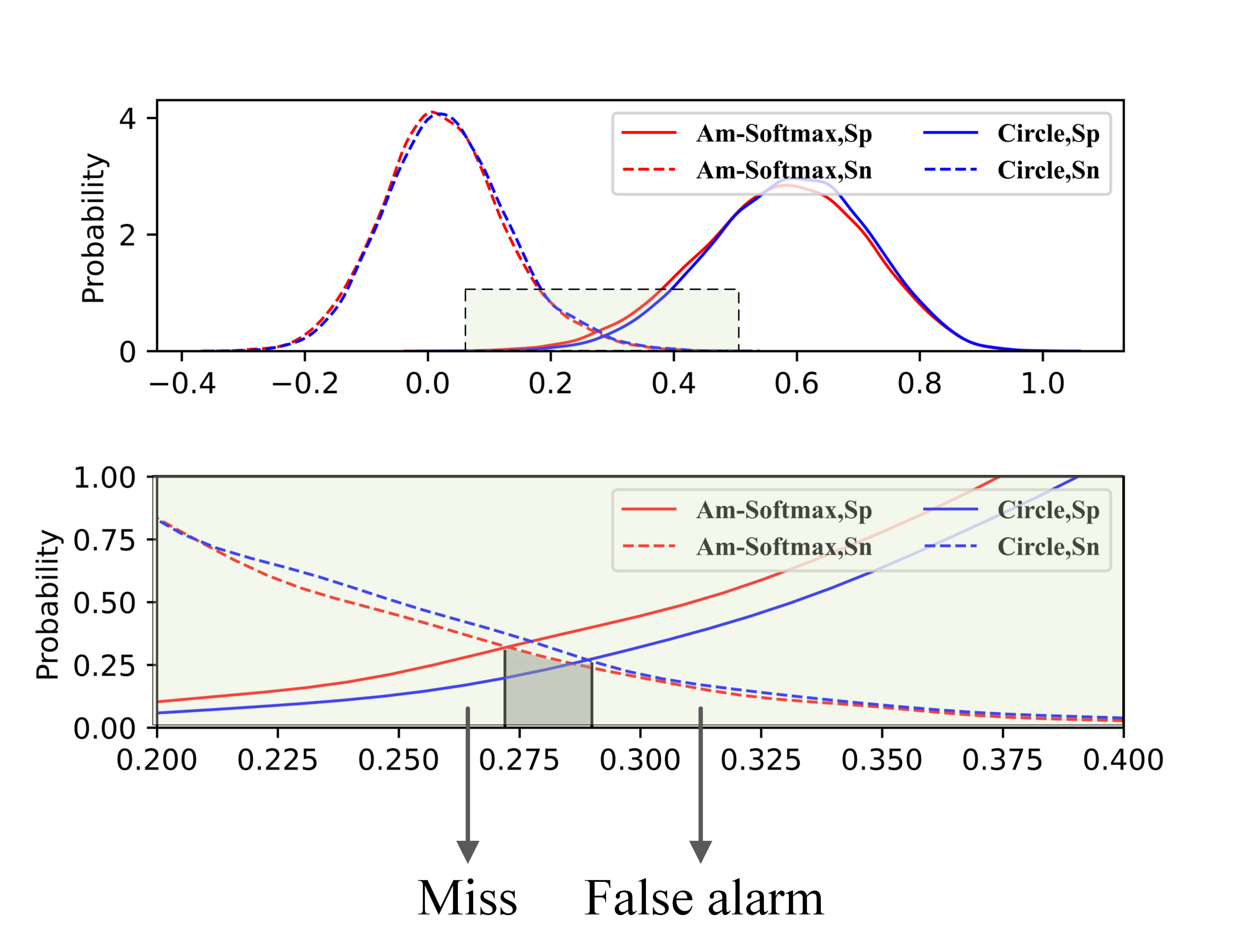}}\\
	\caption{  (a)The similarity distributions during training. (b)The similarity distributions in test set.}
	\label{level}
\end{figure}

\indent We visualized the positive and negative similarity distributions of all training epochs in Figure 3(a) and the similarity distributions of the test set in Figure 3(b). 
In Figure 3(a), the circle loss with $m=0.40$ (the green one) has a lower positive similarity but higher performance than Am-Softmax (the black one), from which we can infer that circle loss with a suitable margin achieves a more reasonable and favorable convergence state. Though the performance of circle loss with $m=0.25$ (the yellow on) is not goodd as $m=0.40$, it achieves the best angle discrimination of training samples, which is better than Am-Softmax. Besides, Circle-Stage (the blue one) gets similar best angle discrimination with $m=0.25$ in the last training stage and maintains good performance in the test set. The performance of Circle-Chunk is slightly worse than Circle-Stage in our experiments. 
\\ \indent As shown in Figure 3(b), the negative similarity distributions of circle loss and Am-Softmax loss are similar, but circle loss has better positive similarity distributions than Am-Softmax loss, especially in the area framed by the dotted line.
The area is enlarged in the figure below, from which we can find that the false alarm of circle loss is smaller than Am-Softmax, and the value is the area of the gray area. So that the EER of circle loss is much lower than Am-Softmax.

\section{Conclusions}
In this paper, we propose a novel angular-based loss called adaptive margin circle loss for speaker verification. Circle loss has flexible optimization and definite convergence status than the other angular-based losses like Am-Softmax. By selecting a fixed appropriate margin, circle loss can achieve promising results. In addition, we explore two strategies, stage-based and chunk-based, to generate a more suitable margin, which can enhance the intra-class compactness of speaker embeddings. With adaptive margin circle loss, our best system achieves 1.31\%EER on Voxceleb1 and 2.13\% on SITW,  core-core, which is competitive among the existing reported results.

\bibliographystyle{IEEEtran}

\bibliography{mybib.bib}

\end{document}